\documentclass[aps,twocolumn,prl,amssymb]{revtex4}
\usepackage{amsmath}
\usepackage{graphicx}
\usepackage{color}
\usepackage{setspace}
\usepackage{multirow}
\usepackage[]{titlesec}
\usepackage{enumerate}

\definecolor{Gray}{gray}{0.9}
\definecolor{LightCyan}{rgb}{0.88,1,1}
\begin{document}
\title{\Large Predicting Bulk Metallic Glass Forming Ability with the Thermodynamic Density of Competing Crystalline States}

\author{Eric Perim}\thanks{These authors have contributed equally to this manuscript.} 
\affiliation{Department of Mechanical Engineering and Materials Science, Duke University, Durham, North Carolina 27708, USA}
\author{Dongwoo Lee}\thanks{These authors have contributed equally to this manuscript.} 
\affiliation{School of Engineering and Applied Sciences, Harvard University, Cambridge, MA 02138, USA}
\author{Yanhui Liu}\thanks{These authors have contributed equally to this manuscript.} 
\affiliation{Department of Mechanical Engineering and Materials Science, Yale University, New Haven, CT 06511, USA}
\author{Cormac Toher}
\affiliation{Department of Mechanical Engineering and Materials Science, Duke University, Durham, North Carolina 27708, USA}
\author{Pan Gong}
\affiliation{Department of Mechanical Engineering and Materials Science, Yale University, New Haven, CT 06511, USA}
\author{Yanglin Li}
\affiliation{Department of Mechanical Engineering and Materials Science, Yale University, New Haven, CT 06511, USA}
\author{W. Neal Simmons}
\affiliation{Department of Mechanical Engineering and Materials Science, Duke University, Durham, North Carolina 27708, USA}
\author{Ohad Levy}
\affiliation{Department of Mechanical Engineering and Materials Science, Duke University, Durham, North Carolina 27708, USA}
\author{Joost Vlassak}
\affiliation{School of Engineering and Applied Sciences, Harvard University, Cambridge, MA 02138, USA}
\author{Jan Schroers}
\affiliation{Department of Mechanical Engineering and Materials Science, Yale University, New Haven, CT 06511, USA}
\author{Stefano Curtarolo}
\email[]{email: stefano@duke.edu}
\affiliation{Materials Science, Electrical Engineering, Physics and Chemistry, Duke University, Durham NC, 27708} 

\date{\today}

\begin{abstract}
  
Despite two decades of studies, the formation of metallic glasses, very promising systems for industrial applications, still remains mostly unexplained. 
This lack of knowledge hinders the search for new systems, still performed with combinatorial trial and error. 
In the past, it was speculated that some sort of {\it ``confusion''} during crystallization could play a key role during their formation. In this article, we propose a
heuristic descriptor quantifying such confusion. It is based on the ``thermodynamic density of competing crystalline states'', parameterized
from high-throughput {\it ab-initio} calculations.
The existence of highly enthalpy-degenerate but geometrically-different phases frustrates the crystallization process and promotes glass formation. 
Two test beds are considered.
A good and a bad glass-former, CuZr and NiZr, are experimentally characterized with high-throughput synthesis. 
The experimental results corroborate the capability of the heuristic descriptor in predicting glass forming ability through the compositional space. 
Our analysis is expected to deepen the understanding of the underlying mechanisms and to accelerate the discovery of novel metallic glasses.

\end{abstract}

\maketitle

Understanding and predicting the formation of multicomponent bulk metallic glasses (BMG) is crucial for fully leveraging their unique
combination of superb mechanical properties \cite{chen2015does} and
plastic-like processability \cite{schroers2006amorphous, Schroers_blow_molding_2011} for potential
applications \cite{Johnson_BMG_2009,Greer_metallic_glasses_review_2009,Schroers_Processing_BMG_2010}. It is also essential for
efficiently identifying the most promising candidates out of the vast propective compositional space \cite{ding2014combinatorial}.
The process underlining the formation of BMGs is still to be fully understood.
It involves a multitude of topological fluctuations competing during
solidification across many length scales \cite{kelton1991crystal, egami2006formation, egami2011atomic}.
Long-range processes, required by the typical non-polymorphic nature of the crystallization, and
atomic-scale fluctuations, precursors of short-range ordered competing phases
\cite{Busch_kinetics_BMG_2007}, are all pitted against each other and against glass formation \cite{kelton1991crystal, kelton2010nucleation, kelton1998new}.
Simulations of amorphous phases have been attempted to disentangle the mechanism of glass formation
\cite{sheng2006atomic,zhang2015glass,zhang2015origin,lee2003criteria,cheng2008relationship,cheng2009atomic,wang2015asymmetric},
within reasonable system sizes, using classical and semi-empirical potentials.
Although they have been successful in investigating the influence of factors such as the atomic size and packing on the glass-forming
ability (GFA), questions about the dynamics of the process still remain, especially considering all the approximations demanded for
performing long molecular dynamics simulations.
Furthermore, adopting {\it ab-initio} methods has been challenging:
even while the most relevant metastable crystalline phases can be calculated and sorted by their energies
\cite{curtarolo:calphad_2005_monster,monsterPGM,curtarolo:art67,curtarolo:art53},
the zero-temperature formalism, lacking vibrational free energy \cite{curtarolo:art65}, and
the absence of an underlying lattice on which to build configurational thermodynamics \cite{deFontaine_ssp_1994,curtarolo:art49}
make the problem impervious to direct computational analysis.

\begin{figure}[h!]
\centerline{\includegraphics[width=0.45\textwidth]{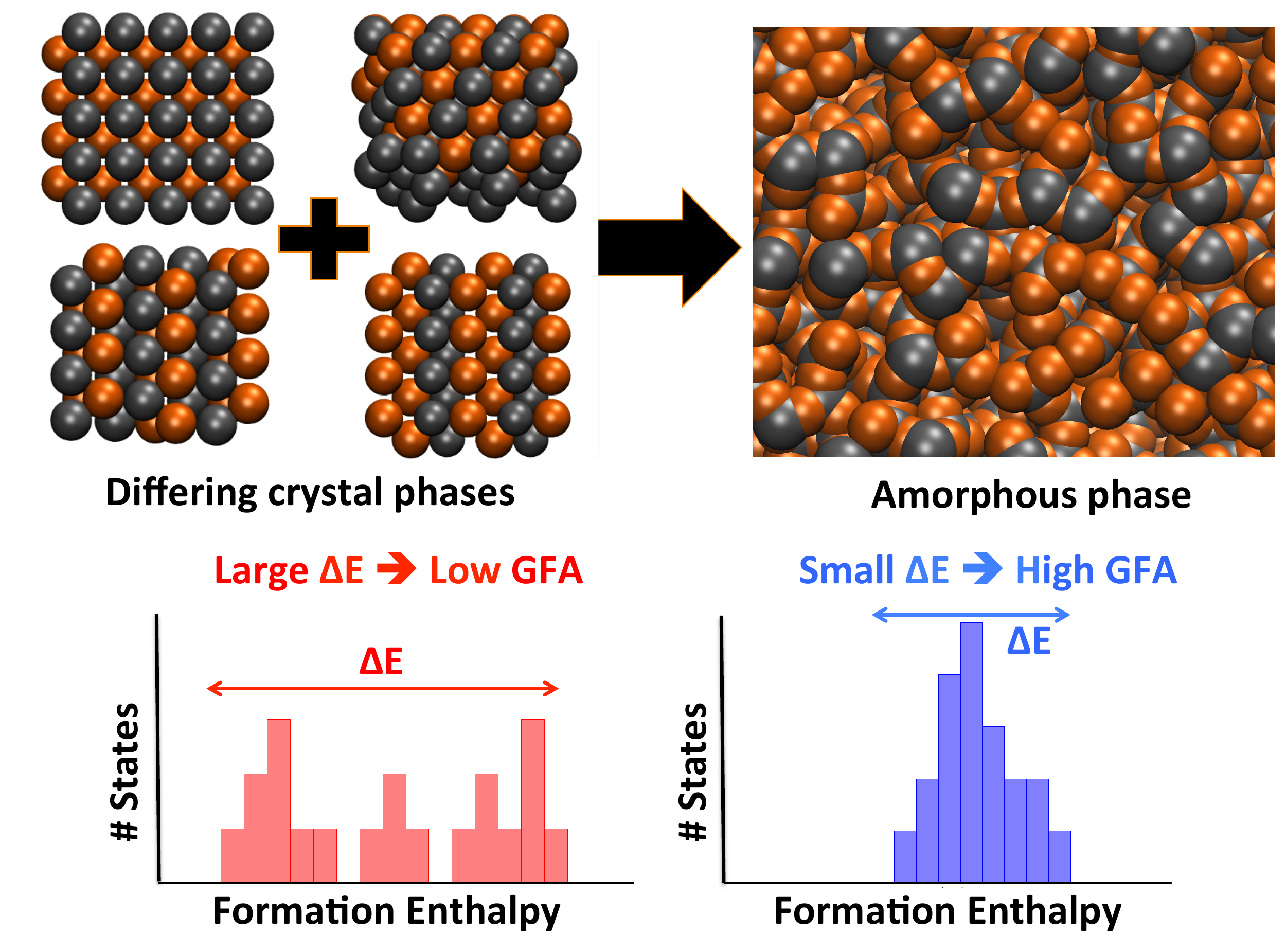}} 
\vspace{-2mm}
\caption{\small 
  Conceptual illustration of the proposed idea. 
  If a particular alloy composition exhibits many structurally different stable and metastable crystal phases which have similar energies, these phases will compete against each other during solidification, disrupting and frustrating the nucleation and crystallization processes and ultimately leading to an amorphous structure.} 
\label{fig1}
\end{figure}

\begin{figure*}[hbt]
\centerline{\includegraphics[width=0.80\textwidth]{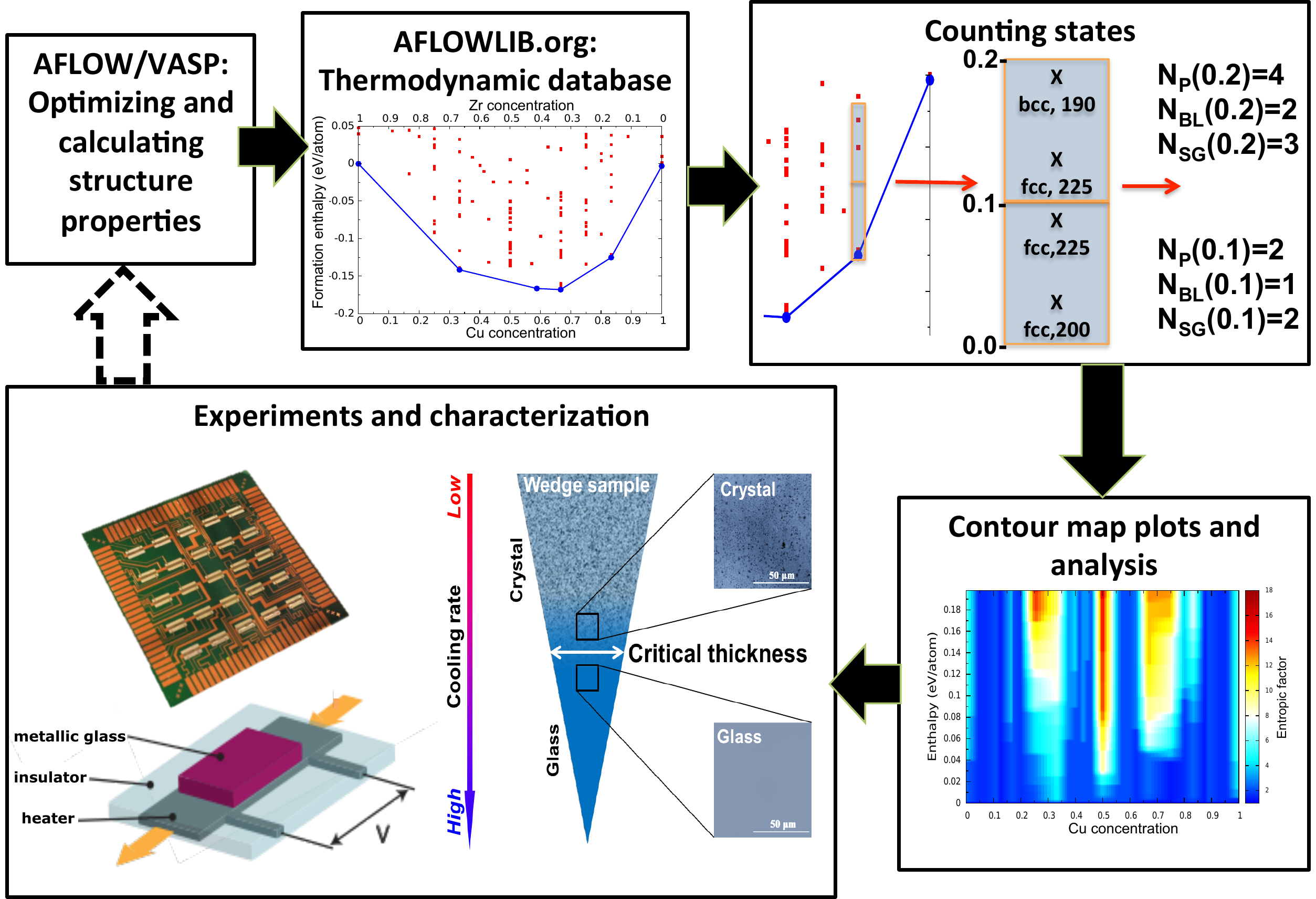}} 
\vspace{-2mm}
\caption{\small 
  Schematic of the adopted approach. 
  Multiple different structures for a given stoichiometry are built using the {\sf AFLOW} prototype libraries \cite{curtarolo:art65}, which are then optimized via VASP calculations adopting the {\sf AFLOW} standard settings \cite{curtarolo:art104}. 
  The resulting data is added to the open thermodynamic database {\sf AFLOWLIB} \cite{curtarolo:art92,aflowlibPAPER}. 
  This data is accessed and used to obtain statistics on the cumulative distribution of entries ($N_P$), Bravais lattices types ($N_{BL}$) and space groups ($N_{SG}$) within a given formation enthalpy range (starting at zero). 
  Contour map plots are created from these distributions, allowing the identification of the best glass forming alloys. Finally, experimental synthesis and characterization are used to verify the computational results.} 
\label{fig2} 
\end{figure*}

Descriptors for bulk glass formation --- correlations between the outcome (glass formation) and other material properties, possibly simpler to characterize \cite{nmatHT} ---
have been proposed based on structural- \cite{miracle2004structural, wang2015asymmetric, cheng2009atomic, cheney2009evaluation}, 
thermodynamic- \cite{cheney2009evaluation, cheney2007prediction, turnbull1969under, busch1995thermodynamics}, 
kinetic- \cite{turnbull1969under,busch1998thermodynamics} and 
electronic structure- considerations \cite{cheng2009atomic, alamgir2000electronic}. 
Nevertheless, a definite and clear picture for predicting GFA still remains to be found.

In a seminal paper \cite{greer1993confusion},
Greer speculated that {\it ``confusion"} during crystallization would promote glass formation.
However, challenges in characterizing such confusion left this direction mostly unexplored.
In this work we tackle this task by a simple straightforward definition of this {\it ``confusion"}.
Our idea is based on the following consideration.
During quenching, crystal growth will occur whenever fluctuations lead to the formation of a crystalline nucleus larger
than a critical size.
Then, in order to obtain an amorphous solid, the formation of critical size nuclei has to be hampered.
We postulate that the existence of multiple phases with very similar energy, implying similar probabilities of being formed,
but dissimilar structures will lead to the formation of several distinct clusters
which will intimately compete and thus keep each other from reaching the critical size needed for crystallization.
Our approach characterizes confusion by the approximate thermodynamic density of distinct structural phases of metastable states,
obtained from \emph{ab-initio} calculations (Figure \ref{fig1}), and concurrent GFA measurements by combinatorial synthesis of alloy libraries and
high-throughput nanocalorimetry.
As test systems, we focus on CuZr and NiZr.
Among the known BMGs, CuZr is probably the most well studied \cite{liu2007cooling,xu2004bulk,wang2004bulk,mei2004binary},
due to its high GFA, easily available components and ease of synthesis.
NiZr, on the other hand, is known for having poor GFA \cite{altounian1983crystallization,fukunaga2006voronoi}.
The contrast between the two glass formers, one strong and one weak, corroborates our ansatz.

\begin{figure*}[htb!]
\centerline{\includegraphics[width=0.95\textwidth]{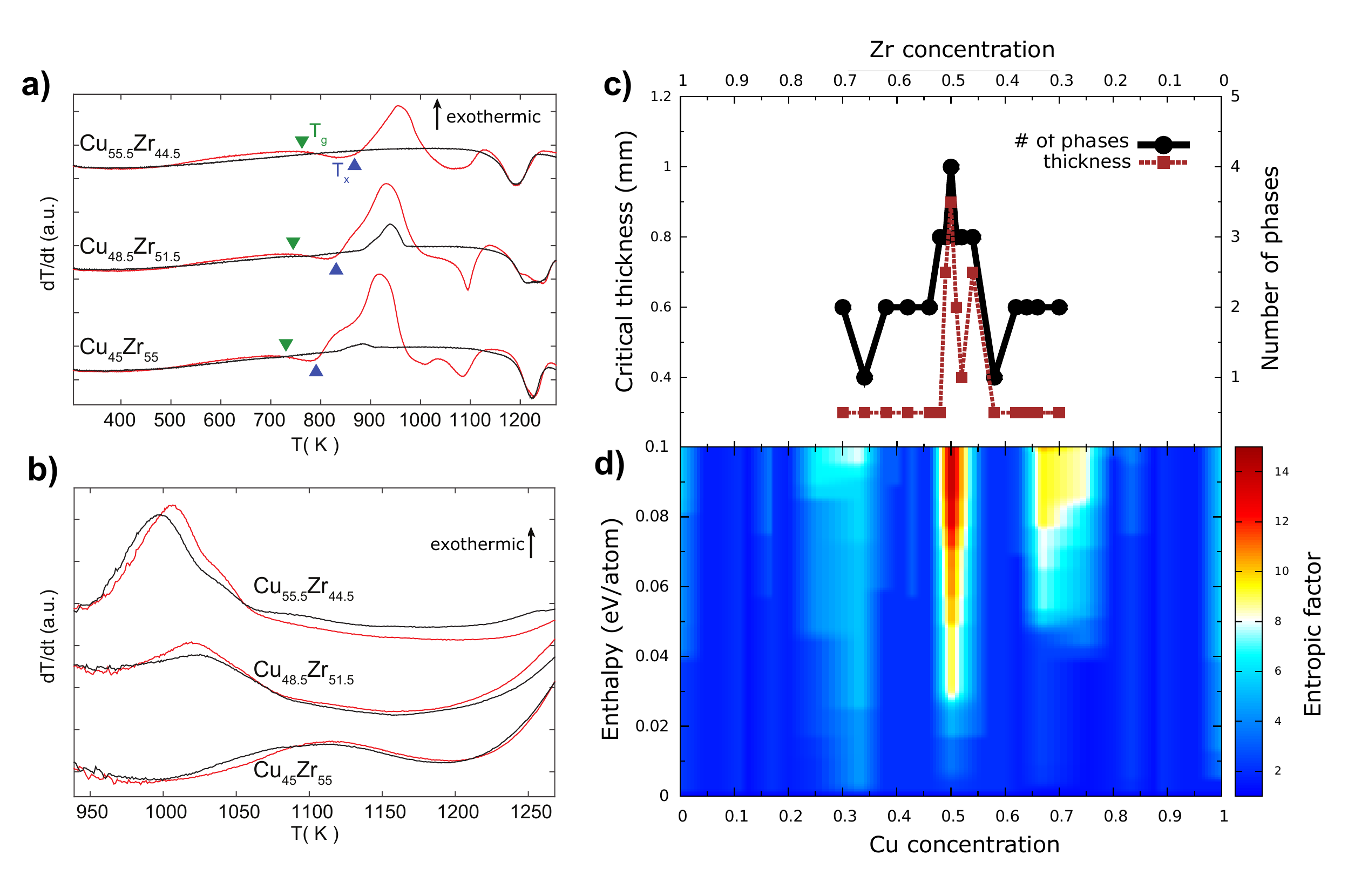}}
\vspace{-6mm}
\caption{\small Nanocalorimetry measurements for CuZr during 
  {\bf (a)} heating and {\bf (b)} cooling at different compositions. The first heating and cooling cycle measurements for the each composition are shown in red, and subsequent measurements are shown in black. 
  {\bf (c)}  Number of phases (solid black line) as measured using XRD, and thickness of the amorphous phase (dashed brown line), determined from the wedge shaped samples, as a function of composition. 
  {\bf (d)} Contour plot of the entropic factor as a function of formation enthalpy (zero corresponds to the ground state of the composition). 
  The color scale represents the entropic factor as calculated using equation \ref{eq1}, for each composition and formation enthalpy difference. 
  This means that, for a given fixed composition ($x$-axis), all phases that are within a given formation enthalpy difference ($y$-axis) from the ground state of that specific composition are used to compute the entropic factor (color scale). 
  Note the sharp peaks both in the number of states observed in experiment and in the entropic factor at the Cu$_{50}$Zr$_{50}$ composition, indicating that the descriptor correctly predicts this composition as having the highest GFA.}
\label{fig3} 
\end{figure*}
Carrying out electronic structure {\it ab-initio} calculations for the infinite number of available states for a given alloy system is obviously impossible, 
especially when no lattice model can be built \cite{deFontaine_ssp_1994,curtarolo:art49}, as in the case of BMGs.
Therefore we adopt the agnostic approach of exploring structural prototypes mostly observed in nature for these type of systems.
The method, shown to be capable of reasonably sampling the phase space and predicting novel compounds \cite{curtarolo:calphad_2005_monster,monsterPGM,curtarolo:art67,curtarolo:art53,curtarolo:art49},
is expected to estimate the thermodynamic density of states for an alloy system. 
We use the binary alloy data available in the {\sf AFLOWLIB} set of repositories \cite{curtarolo:art92,aflowlibPAPER} to count the number of different structural phases 
in a given formation enthalpy range as a function of the composition. 
All calculations available in the {\sf AFLOWLIB} database are done utilizing the {\sf VASP} \cite{kresse_vasp, vasp, vasp_cms1996} code within the {\sf AFLOW} 
computational materials design framework \cite{curtarolo:art65,curtarolo:art104}, at the density functional theory level of approximation. 
The binary alloy systems are fully relaxed in accordance with the {\sf AFLOW} standard settings \cite{curtarolo:art104}; 
which uses the GGA-PBE \cite{PBE, PBE2} exchange-correlation, PAW potentials \cite{PAW, kresse_vasp_paw}, 
at least 6000 {\bf k}-points per reciprocal atom (KPPRA) and a plane wave cut-off at least 1.4 times the largest value recommended for the {\sf VASP} potentials of the constituents.
The multiple different crystalline phases for each particular stoichiometry are built from the {\sf AFLOW} library of common prototypes \cite{curtarolo:art65}.

To quantify the level of disorder associated with an alloy system, we proceed by identifying the groundstates and then
counting all of the available phases at the corresponding compositions,
ordered by their formation enthalpy difference above the respective ground state, $\Delta H\equiv H-H_{gs}$,
effectively building a cumulative distribution of the number of phases, $N_P(H-H_{gs})$ (see Figure \ref{fig2}).
We also count the number of different Bravais lattice types, $N_{BL}(\Delta H)$, and space groups, $N_{SG}(\Delta H)$. 
We heuristically combine these three quantities into a single descriptor, called {\it ``entropic factor'}', $S_F(\Delta H)$, 
defined as the cubic root of their product: 
\begin{equation} 
  S_F(\Delta H)=\sqrt[3]{N_P(\Delta H)\times N_{BL}(\Delta H)\times N_{SG}(\Delta H)}.
  \label{eq1} 
\end{equation}
$S_F(\Delta H)$ is related to the configurational entropy at a given composition but, by taking into account the different
symmetries available to the system, it is more generally representative
of the frustration towards the crystallization of a single homogeneous crystal structure.
Compositions with large $S_F(\Delta H)$ are expected to present structures with more disorder, thus leading to high GFA.
In this analysis, the formation enthalpies, Bravais lattices and space groups were determined
from the calculated energies and symmetries of the relaxed relevant structures.

X-ray diffraction (XRD) and scanning electron microscopy (SEM) measurements were performed on ingots of CuZr and NiZr alloys prepared by arc-melting the pure elements under an argon atmosphere. 
The alloys were re-melted and suction cast into a wedge-shaped cavity in a copper mold. 
The as-cast rods were cut into half along the longitudinal direction and polished to a mirror finish followed by etching. 
GFA was evaluated by observing under scanning electron microscope the contrast change along the longitudinal direction. 
The critical thickness was determined when the transition from featureless contrast to clear microstructure could be observed, as shown in Figure \ref{fig2}. 
The crystalline and amorphous structures were further identified by XRD using a Cu-$K_\alpha$ source. 

We also synthesized and characterized thin-film samples deposited by magnetron sputtering elementary targets (99.99\% pure) inside a vacuum chamber with a base pressure better than $ 2\times10^{-7}$ Torr. Sputter deposition results in an effective quenching rate greater than $10^{10}$ K/s, allowing a broad range of alloys to be obtained in the amorphous state.

Nanocalorimetry measurements were performed on thin-film samples of the binary alloys using micromachined calorimetry sensors \cite{mccluskey2010combinatorial,mccluskey2010nano,lee2013scanning,lee2014low}. 
The measurements were performed in vacuum at nominal heating rates ranging from 2000 to 8500 K/s, and cooling rates of approximately 5000 K/s. 
All samples were repeatedly heated to 1300 K to evaluate the crystallization behavior both in the as-deposited state and after melt/quenching. 
Nanocalorimetry measurements reveal the glass transition, crystallization and liquidus temperatures. These quantities allow us to estimate GFA.

\begin{figure}[htb!] 
  \centerline{\includegraphics[width=0.49\textwidth]{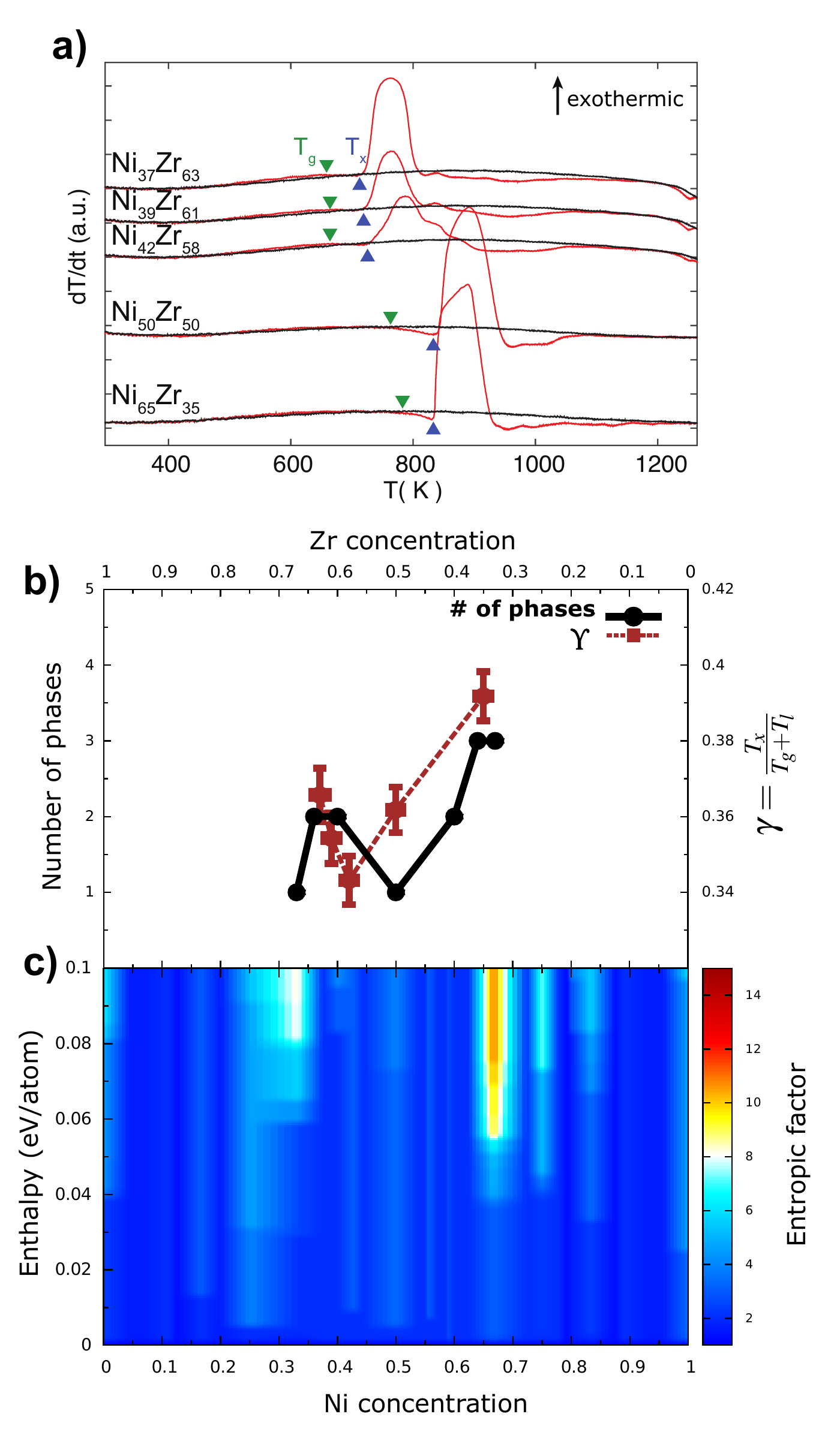}}
  \vspace{-6mm}
  \caption{\small 
    {\bf (a)} Nanocalorimetry measurements for NiZr during heating at different compositions. 
    The first heating cycle measurements for each composition are shown in red, and subsequent measurements are shown in black.
    {\bf (b)} Number of phases (solid black line) as measured using XRD, and $\gamma$ descriptor calculated for NiZr alloys (dashed brown line). 
    {\bf (c)} Contour plot of the entropic factor as a function of formation enthalpy (zero corresponds to the ground state of the composition). 
    Note the sharp peaks both in the number of states observed in experiment and in the entropic factor at the Ni$_{35}$Zr$_{65}$ and Ni$_{65}$Zr$_{35}$ compositions.} 
  \label{fig4}
\end{figure}

Figures \ref{fig3}(a) and (b) show the nanocalorimetry results for the CuZr binary alloy with compositions in the bulk glass forming region. 
Each measurement consisted of two thermal cycles in which the thin-film samples were heated to above the melting point and then quenched. 
All samples show clear signals corresponding to glass transition, crystallization, and melting when first heated from the as-deposited state, indicating that they were deposited in the amorphous state, Figure \ref{fig3}(a).
A better glass former has a lower critical cooling rate, so the amount of amorphous phase recovered after melt/quenching should scale with GFA. 
We observe in Figure \ref{fig3}(a) that the magnitude of the crystallization peak after the first thermal cycle changes significantly with composition: Cu$_{48.5}$Zr$_{51.5}$ has the strongest crystallization peak 
and is thus expected to have the highest GFA among the samples tested; Cu$_{55.5}$Zr$_{44.5}$ on the other hand has no discernable crystallization peak. 
This result is confirmed by calorimetry measurements obtained after cooling from the melted state: 
the heat released upon solidification results in an exothermic peak in the cooling curve; 
the magnitude of this peak scales with the amount of crystalline phase formed on quenching and should be inversely proportional to the GFA, Figure \ref{fig3}(b). 
The experimentally observed number of phases and the amorphous phase thickness obtained from the casting experiments are shown in Figure \ref{fig3}(c). 
The calculated entropic factor, Figure \ref{fig3}(d), can be compared with these two quantities, and the results show very good agreement between all methods that Cu$_{50}$Zr$_{50}$ is the best glass forming composition.

Figure \ref{fig4} shows similar measurements for the NiZr alloy system, which has been shown to be a weak glass former \cite{altounian1983crystallization,fukunaga2006voronoi}. 
Although as-deposited samples were amorphous and showed distinct crystallization peaks, subsequent melt/quenching did not produce any amorphous samples,
and no crystallization peaks are observed in scans obtained after melting (Figure \ref{fig4}(a)). Instead, we use $\gamma = {T_x}/{(T_g+T_l)}$, defined in Ref. \onlinecite{lu2002new} and shown in Figure \ref{fig4}(b), 
as a less direct measure of GFA. Figures \ref{fig4}(b) and (c) show strong correlation between the experimental measurements and the entropic factor descriptor.
There is a very weak GFA peak around Ni$_{35}$Zr$_{65}$ according to the experimental measurements, which is predicted to be around Ni$_{30}$Zr$_{70}$ by the entropic factor descriptor. 
A more pronounced peak is measured around Ni$_{65}$Zr$_{35}$, which is also successfully predicted in the same region by the entropic factor descriptor. 
Thus, the new proposed predictor correlates well with the traditional empirical indicators of glass formation in metallic alloys, with an accuracy of the order of $5\%$ in  composition, which is quite satisfactory.
In addition, comparing Figures \ref{fig3}(c) and \ref{fig4}(c), it is clear that the entropic facor exhibited by the high GFA alloy CuZr is significantly higher
than that shown by the low GFA NiZr alloy, thus correctly pointing out the more favorable alloy system for glass formation.

In conclusion, in this article we propose a simple computational descriptor for GFA. It is based on the number and energy features
of distinct but competing crystalline phases near the ground state of a given metallic alloy.
The descriptor successfully differentiates the two experimentally studied test cases: a good and a bad glass-former,
CuZr and NiZr respectively.
It thus exhibits the capability of providing predictions of glass formation potential from first principles computer
calculations, which would allow
large scale high-throughput screening of the very large composition space available for glass forming alloys.
Such large scale computational studies
should reveal correlations that will contribute to a better understanding of the physics underlying glass formation, as well as accelerate the development of new BMGs.

This work was supported by the National Science Foundation under DMREF Grants No. DMR-1436151, DMR-1436268, and DMR-1435820.
Experiments were performed in part at the Center for Nanoscale Systems at Harvard University (supported by NSF under Award No. ECS-0335765), 
at the Materials Research Science and Engineering Center at Harvard University (supported by NSF under Award No. DMR-1420570), and
at the Materials Research Science and Engineering Center at Yale University (supported by NSF under Award No. DMR-1119826).
Calculations were performed at the Duke University - Center for Materials Genomics.
E.P., C.T. and S.C. acknowledge partial support by the Department of Defense - Office of Naval Research (N00014-13-1-0635, N00014-14-1-0526).


\begin{thebibliography}{10}

\bibitem{chen2015does}
W.~Chen, J.~Ketkaew, Z.~Liu, R.~M. {Ojeda~Mota}, K.~O'Brien, C.~S. {da~Silva},
  and J.~Schroers, \emph{Does the fracture toughness of bulk metallic glasses
  scatter?}, Scr.\ Mater. \textbf{107}, 1--4 (2015).

\bibitem{schroers2006amorphous}
J.~Schroers and N.~Paton, \emph{Amorphous metal alloys form like plastics},
  Advanced Materials \& Processes p.~61 (2006).

\bibitem{Schroers_blow_molding_2011}
J.~Schroers, T.~M. Hodges, G.~Kumar, H.~Raman, A.~J. Barnes, Q.~Pham, and T.~A.
  Waniuk, \emph{Thermoplastic blow molding of metals}, Materials Today
  \textbf{14}, 14--19 (2011).

\bibitem{Johnson_BMG_2009}
W.~L. Johnson, \emph{Bulk Glass-Forming Metallic Alloys: Science and
  Technology}, MRS\ Bull. \textbf{24}, 42--56 (1999).

\bibitem{Greer_metallic_glasses_review_2009}
A.~L. Greer, \emph{Metallic glasses...on the threshold}, Materials Today
  \textbf{12}, 14--22 (2009).

\bibitem{Schroers_Processing_BMG_2010}
J.~Schroers, \emph{Processing of Bulk Metallic Glass}, Adv.\ Mater.
  \textbf{22}, 1566--1597 (2010).

\bibitem{ding2014combinatorial}
S.~Ding, Y.~Liu, Y.~Li, Z.~Liu, S.~Sohn, F.~J. Walker, and J.~Schroers,
  \emph{Combinatorial development of bulk metallic glasses}, Nature\ Mater.
  \textbf{13}, 494--500 (2014).

\bibitem{kelton1991crystal}
K.~F. Kelton, \emph{Crystal nucleation in liquids and glasses}, Solid\ State\
  Phys. \textbf{45}, 75--177 (1991).

\bibitem{egami2006formation}
T.~Egami, \emph{Formation and deformation of metallic glasses: atomistic
  theory}, Intermetallics \textbf{14}, 882--887 (2006).

\bibitem{egami2011atomic}
T.~Egami, \emph{Atomic level stresses}, Prog.\ Mater.\ Sci. \textbf{56},
  637--653 (2011).

\bibitem{Busch_kinetics_BMG_2007}
R.~Busch, J.~Schroers, and W.~H. Wang, \emph{Thermodynamics and Kinetics of
  Bulk Metallic Glass}, MRS\ Bull. \textbf{32}, 620--623 (2007).

\bibitem{kelton2010nucleation}
K.~Kelton and A.~L. Greer, \emph{Nucleation in condensed matter: applications
  in materials and biology}, vol.~15 (Elsevier, 2010).

\bibitem{kelton1998new}
K.~F. Kelton, \emph{A new model for nucleation in bulk metallic glasses}
  \textbf{77}, 337--344 (1998).

\bibitem{sheng2006atomic}
H.~W. Sheng, W.~K. Luo, F.~M. Alamgir, J.~M. Bai, and E.~Ma, \emph{Atomic
  packing and short-to-medium-range order in metallic glasses}, Naturere
  \textbf{439}, 419--425 (2006).

\bibitem{zhang2015glass}
K.~Zhang, Y.~Liu, J.~Schroers, M.~D. Shattuck, and C.~S. O'Hern, \emph{The
  glass-forming ability of model metal-metalloid alloys}, J.\ Chem.\ Phys.
  \textbf{142}, 104504 (2015).

\bibitem{zhang2015origin}
K.~Kai, B.~Dice, Y.~Liu, J.~Schroers, M.~D. Shattuck, and C.~S. O'Hern,
  \emph{On the origin of multi-component bulk metallic glasses: Atomic size
  mismatches and de-mixing}, J.\ Chem.\ Phys. \textbf{143}, 054501 (2015).

\bibitem{lee2003criteria}
H.-J. Lee, T.~Cagin, W.~L. Johnson, and W.~A. Goddard~III, \emph{Criteria for
  formation of metallic glasses: The role of atomic size ratio}, J.\ Chem.\
  Phys. \textbf{119}, 9858--9870 (2003).

\bibitem{cheng2008relationship}
Y.~Q. Cheng, H.~W. Sheng, and E.~Ma, \emph{Relationship between structure,
  dynamics, and mechanical properties in metallic glass-forming alloys}, Phys.\
  Rev.\ B \textbf{78}, 014207 (2008).

\bibitem{cheng2009atomic}
Y.~Q. Cheng, E.~Ma, and H.~W. Sheng, \emph{Atomic Level Structure in
  Multicomponent Bulk Metallic Glass}, Phys.\ Rev.\ Lett. \textbf{102}, 245501
  (2009).

\bibitem{wang2015asymmetric}
M.~Wang, K.~Zhang, Z.~Li, Y.~Liu, J.~Schroers, M.~D. Shattuck, and C.~S.
  O'Hern, \emph{Asymmetric crystallization during cooling and heating in model
  glass-forming systems}, Phys.\ Rev.\ E \textbf{91}, 032309 (2015).

\bibitem{curtarolo:calphad_2005_monster}
S.~Curtarolo, D.~Morgan, and G.~Ceder, \emph{Accuracy of {\it ab initio}
  methods in predicting the crystal structures of metals: {A} review of 80
  binary alloys}, Calphad \textbf{29}, 163--211 (2005).

\bibitem{monsterPGM}
G.~L.~W. Hart, S.~Curtarolo, T.~B. Massalski, and O.~Levy, \emph{Comprehensive
  Search for New Phases and Compounds in Binary Alloy Systems Based on
  {Platinum}-Group Metals, Using a Computational First-Principles Approach},
  Phys.\ Rev.\ X \textbf{3}, 041035 (2013).

\bibitem{curtarolo:art67}
M.~Jahn\'{a}tek, O.~Levy, G.~L.~W. Hart, L.~J. Nelson, R.~V. Chepulskii,
  J.~Xue, and S.~Curtarolo, \emph{Ordered phases in ruthenium binary alloys
  from high-throughput first-principles calculations}, Phys.\ Rev.\ B
  \textbf{84}, 214110 (2011).

\bibitem{curtarolo:art53}
O.~Levy, R.~V. Chepulskii, G.~L.~W. Hart, and S.~Curtarolo, \emph{The New face
  of {Rhodium} Alloys: Revealing Ordered Structures from First Principles}, J.\
  Am.\ Chem.\ Soc. \textbf{132}, 833--837 (2010).

\bibitem{curtarolo:art65}
S.~Curtarolo, W.~Setyawan, G.~L.~W. Hart, M.~Jahn\'{a}tek, R.~V. Chepulskii,
  R.~H. Taylor, S.~Wang, J.~Xue, K.~Yang, O.~Levy, M.~J. Mehl, H.~T. Stokes,
  D.~O. Demchenko, and D.~Morgan, \emph{{AFLOW}: An automatic framework for
  high-throughput materials discovery}, Comp.\ Mat.\ Sci. \textbf{58}, 218--226
  (2012).

\bibitem{deFontaine_ssp_1994}
D.~de~Fontaine, \emph{Cluster Approach to Order-Disorder Transformations in
  Alloys}, in \emph{Solid State Physics}, edited by H.~Ehrenreich and
  D.~Turnbull (Wiley, New York, 1994), vol.~47, pp. 33--176.

\bibitem{curtarolo:art49}
O.~Levy, G.~L.~W. Hart, and S.~Curtarolo, \emph{Uncovering Compounds by Synergy
  of Cluster Expansion and High-Throughput Methods}, J.\ Am.\ Chem.\ Soc.
  \textbf{132}, 4830--4833 (2010).

\bibitem{curtarolo:art104}
C.~E. Calderon, J.~J. Plata, C.~Toher, C.~Oses, O.~Levy, M.~Fornari, A.~Naturen,
  M.~J. Mehl, G.~L.~W. Hart, M.~{Buongiorno~Nardelli}, and S.~Curtarolo,
  \emph{The {AFLOW} Standard for High-Throughput Materials Science
  Calculations}, Comp.\ Mat.\ Sci. \textbf{108 Part A}, 233--238 (2015).

\bibitem{curtarolo:art92}
R.~H. Taylor, F.~Rose, C.~Toher, O.~Levy, K.~Yang, M.~{Buongiorno~Nardelli},
  and S.~Curtarolo, \emph{A {RESTful API} for exchanging Materials Data in the
  {AFLOWLIB.org} consortium}, Comp.\ Mat.\ Sci. \textbf{93}, 178--192 (2014).

\bibitem{aflowlibPAPER}
S.~Curtarolo, W.~Setyawan, S.~Wang, J.~Xue, K.~Yang, R.~H. Taylor, L.~J.
  Nelson, G.~L.~W. Hart, S.~Sanvito, M.~{Buongiorno~Nardelli}, N.~Mingo, and
  O.~Levy, \emph{{AFLOWLIB.ORG}: A distributed materials properties repository
  from high-throughput {\it ab initio} calculations}, Comp.\ Mat.\ Sci.
  \textbf{58}, 227--235 (2012).

\bibitem{nmatHT}
S.~Curtarolo, G.~L.~W. Hart, M.~{Buongiorno~Nardelli}, N.~Mingo, S.~Sanvito,
  and O.~Levy, \emph{The high-throughput highway to computational materials
  design}, Nature\ Mater. \textbf{12}, 191--201 (2013).

\bibitem{miracle2004structural}
D.~B. Miracle, \emph{A structural model for metallic glasses}, Nature\ Mater.
  \textbf{3}, 697--702 (2004).

\bibitem{cheney2009evaluation}
J.~Cheney and K.~Vecchio, \emph{Evaluation of glass-forming ability in metals
  using multi-model techniques}, J.\ Alloys Compound. \textbf{471}, 222--240
  (2009).

\bibitem{cheney2007prediction}
J.~Cheney and K.~Vecchio, \emph{Prediction of glass-forming compositions using
  liquidus temperature calculations}, Materials Science and Engineering: A
  \textbf{471}, 135--143 (2007).

\bibitem{turnbull1969under}
D.~Turnbull, \emph{Under what conditions can a glass be formed?} \textbf{10},
  473--488 (1969).

\bibitem{busch1995thermodynamics}
R.~Busch, Y.~J. Kim, and W.~L. Johnson, \emph{Thermodynamics and kinetics of
  the undercooled liquid and the glass transition of the
  Zr$_{41.2}$Ti$_{13.8}$Cu$_{12.5}$Ni$_{10.0}$Be$_{22.5}$ alloy}, J.\ Appl.\
  Phys. \textbf{77}, 4039--4043 (1995).

\bibitem{busch1998thermodynamics}
R.~Busch, W.~Liu, and W.~L. Johnson, \emph{Thermodynamics and kinetics of the
  Mg$_{65}$Cu$_{25}$Y$_{10}$ bulk metallic glass forming liquid}, J.\ Appl.\
  Phys. \textbf{83}, 4134--4141 (1998).

\bibitem{alamgir2000electronic}
F.~M. Alamgir, H.~Jain, R.~B. Schwarz, O.~Jin, and D.~B. Williams,
  \emph{Electronic structure of Pd-based bulk metallic glasses} \textbf{274},
  289--293 (2000).

\bibitem{greer1993confusion}
A.~L. Greer, \emph{Confusion by design}, Naturere \textbf{366}, 303--304 (1993).

\bibitem{liu2007cooling}
Y.~Liu, H.~Bei, C.~T. Liu, and E.~P. George, \emph{Cooling-rate induced
  softening in a Zr$_{50}$Cu$_{50}$ bulk metallic glass}, Appl.\ Phys.\ Lett.
  \textbf{90}, 071909 (2007).

\bibitem{xu2004bulk}
D.~Xu, B.~Lohwongwatana, G.~Duan, W.~L. Johnson, and C.~Garland, \emph{Bulk
  metallic glass formation in binary Cu-rich alloy series-Cu$_{100-x}$Zr$_x$
  (x= 34, 36, 38.2, 40 at.\%) and mechanical properties of bulk
  Cu$_{64}$Zr$_{36}$ glass}, Acta\ Mater. \textbf{52}, 2621--2624 (2004).

\bibitem{wang2004bulk}
D.~Wang, Y.~Li, B.~B. Sun, M.~L. Sui, K.~Lu, and E.~Ma, \emph{Bulk metallic
  glass formation in the binary Cu-Zr system}, Appl.\ Phys.\ Lett. \textbf{84},
  4029--4031 (2004).

\bibitem{mei2004binary}
T.~Mei-Bo, Z.~De-Qian, P.~Ming-Xiang, and W.~Wei-Hua, \emph{Binary Cu-Zr bulk
  metallic glasses}, Chinese Physics Letters \textbf{21}, 901 (2004).

\bibitem{altounian1983crystallization}
Z.~Altounian, T.~Guo-hua, and J.~O. Strom-Olsen, \emph{Crystallization
  characteristics of Ni-Zr metallic glasses from Ni$_{20}$Zr$_{80}$ to
  Ni$_{70}$Zr$_{30}$}, J.\ Appl.\ Phys. \textbf{54}, 3111--3116 (1983).

\bibitem{fukunaga2006voronoi}
T.~Fukunaga, K.~Itoh, T.~Otomo, K.~Mori, M.~Sugiyama, H.~Kato, M.~Hasegawa,
  A.~Hirata, Y.~Hirotsu, and A.~C. Hannon, \emph{Voronoi analysis of the
  structure of Cu-Zr and Ni-Zr metallic glasses}, Intermetallics \textbf{14},
  893--897 (2006).

\bibitem{kresse_vasp}
G.~Kresse and J.~Hafner, \emph{{\it Ab initio} molecular dynamics for liquid
  metals}, Phys.\ Rev.\ B \textbf{47}, 558--561 (1993).

\bibitem{vasp}
G.~Kresse and J.~Furthm\"uller, \emph{Efficient iterative schemes for {\it ab
  initio} total-energy calculations using a plane-wave basis set}, Phys.\ Rev.\
  B \textbf{54}, 11169--11186 (1996).

\bibitem{vasp_cms1996}
G.~Kresse and J.~Furthm\"uller, \emph{Efficiency of ab-initio total energy
  calculations for metals and semiconductors using a plane-wave basis set},
  Comp.\ Mat.\ Sci. \textbf{6}, 15 (1996).

\bibitem{PBE}
J.~P. Perdew, K.~Burke, and M.~Ernzerhof, \emph{Generalized gradient
  approximation made simple}, Phys.\ Rev.\ Lett. \textbf{77}, 3865--3868
  (1996).

\bibitem{PBE2}
J.~P. Perdew, K.~Burke, and M.~Ernzerhof, \emph{Erratum: Generalized Gradient
  Approximation Made Simple}, Phys.\ Rev.\ Lett. \textbf{78}, 1396 (1997).

\bibitem{PAW}
P.~E. Bl\"ochl, \emph{Projector augmented-wave method}, Phys.\ Rev.\ B
  \textbf{50}, 17953--17979 (1994).

\bibitem{kresse_vasp_paw}
G.~Kresse and D.~Joubert, \emph{From ultrasoft pseudopotentials to the
  projector augmented-wave method}, Phys.\ Rev.\ B \textbf{59}, 1758 (1999).

\bibitem{mccluskey2010combinatorial}
P.~J. McCluskey and J.~J. Vlassak, \emph{Combinatorial nanocalorimetry}, J.\
  Mater.\ Res. \textbf{25}, 2086--2100 (2010).

\bibitem{mccluskey2010nano}
P.~J. McCluskey and J.~J. Vlassak, \emph{Nano-thermal transport array: An
  instrument for combinatorial measurements of heat transfer in nanoscale
  films}, Thin Solid Films \textbf{518}, 7093--7106 (2010).

\bibitem{lee2013scanning}
D.~Lee, G.-D. Sim, K.~Xiao, Y.~S. Choi, and J.~J. Vlassak, \emph{Scanning AC
  nanocalorimetry study of Zr/B reactive multilayers}, J.\ Appl.\ Phys.
  \textbf{114}, 214902 (2013).

\bibitem{lee2014low}
D.~Lee, G.-D. Sim, K.~Xiao, and J.~J. Vlassak, \emph{Low-Temperature Synthesis
  of Ultra-High-Temperature Coatings of ZrB2 Using Reactive Multilayers}, J.\
  Phys.\ Chem.\ C \textbf{118}, 21192--21198 (2014).

\bibitem{lu2002new}
Z.~P. Lu and C.~T. Liu, \emph{A new glass-forming ability criterion for bulk
  metallic glasses}, Acta\ Mater. \textbf{50}, 3501--3512 (2002).

\end{thebibliography}
\end{document}